\documentclass[reprint,superscriptaddress,amsmath,amssymb,amsfonts,aps,prl,floatfix]{revtex4-2}
\usepackage{times}
\usepackage{graphicx}
\usepackage{dcolumn}
\usepackage{bm}
\usepackage{ulem}
\usepackage[colorlinks=true, citecolor=blue, linkcolor=blue, urlcolor=blue]{hyperref}
\usepackage{bbold}
\DeclareMathAlphabet\mathbfcal{OMS}{cmsy}{b}{n}
\usepackage{wasysym}
\usepackage[usenames,dvipsnames]{xcolor}

\begin{document}


\title{Strain Disorder and Gapless Intervalley Coherent Phase in Twisted Bilayer Graphene
}
\author{Gal Shavit}
\affiliation{Department of Condensed Matter Physics,    Weizmann Institute of Science, Rehovot, Israel 7610001}

\author{Kry\v{s}tof Kol\'{a}\v{r}}
\affiliation{\mbox{Dahlem Center for Complex Quantum Systems and Fachbereich Physik, Freie Universit\"at Berlin, 14195 Berlin, Germany}}

\author{Christophe Mora}
\affiliation{Universit\'e Paris Cit\'e, CNRS,  Laboratoire  Mat\'eriaux  et  Ph\'enom\`enes  Quantiques, 75013  Paris,  France}

\author{Felix von Oppen}
\affiliation{\mbox{Dahlem Center for Complex Quantum Systems and Fachbereich Physik, Freie Universit\"at Berlin, 14195 Berlin, Germany}}

\author{Yuval Oreg}
\affiliation{Department of Condensed Matter Physics,    Weizmann Institute of Science, Rehovot, Israel 7610001}

\date{\today}

\begin{abstract}
Correlated insulators are frequently observed in magic angle twisted bilayer graphene at even fillings of electrons or holes per moir\'e unit-cell.
Whereas theory predicts these insulators to be intervalley coherent excitonic phases, the measured gaps are routinely much smaller than theoretical estimates.
We explore the effects of random strain variations on the intervalley coherent phase, which have a pair-breaking effect analogous to magnetic disorder in superconductors.
We find that the spectral gap may be strongly suppressed by strain disorder, or vanish altogether, even as intervalley coherence is maintained.
We discuss predicted features of the tunneling density of states, show that the activation gap measured in transport experiments corresponds to the diminished gap, and thus offer a solution for the apparent discrepancy between the theoretical and experimental gaps.
\end{abstract}

\maketitle

\textit{Introduction.---}In recent years, magic-angle twisted bilayer graphene (MATBG)~\cite{BistritzermacdonaldPNAS} has emerged as an exciting and versatile platform for strong-correlation physics.
Its rich phase diagram prominently features correlated insulators, Chern insulators, superconductivity, and strange metallicity~\cite{CaoCorrelatedInsulator,CaoUnconventionalSC,EfetovAllIntegers,YankowitzTuningMATBG,YoungTuningSC,DiracRevivals,YazdaniRevivals,BLGscreening,hBNgoldhaberGordon,hBNyoung,EfetovBernevigPRLChernSC}.

Correlated insulators consistently found at fillings $\nu=\pm 2$ electrons per moir\'e unit cell relative to charge neutrality~(CN), and more rarely at CN, have been suggested to originate from Kramers intervalley-coherent order (K-IVC)~\cite{BultnickKhalaf2020,hofmann2021fermionic,ShavitMaATBGprl,OxfordMATBG}.
The K-IVC phase is in some sense analogous to a superconductor, with the condensate made up of intervalley electron-holes pairs.
Interestingly, theoretical predictions of the K-IVC gap energy, by both Hartree-Fock and numerically exact methods, overestimate the measured activation gap of $\sim{\cal O}\left(1\,{\rm meV}\right)$~\cite{CaoCorrelatedInsulator,EfetovAllIntegers,YankowitzTuningMATBG,YoungTuningSC,BLGscreening} by more than an order of magnitude.

It has been experimentally well-established that the moir\'e lattice formed in realistic MATBG devices is not pristine, presumably due to substantial relaxation effects of the underlying graphene lattice \cite{KoshinoStrain}.
Disorder in the local twist angle between the graphene layers has been observed~\cite{Uri2020Zeldovtwistangledisorder,DiracRevivals}, leading to domains with slightly different effective moir\'e unit-cell sizes.
Local measurements have also shown significant strain effects, consistent with a moir\'e lattice distortion of $0.1\%-0.7\%$~\cite{YazdaniStrain2019,CoryDeanStrain2019,NadjPergeStrain2019}.

As the two graphene layers may be subjected to different strain fields,
the strain tensor applied to the bilayer is comprised of a layer-symmetric part (homostrain) and a layer-antisymmetric contribution (heterostrain).
Uniform heterostrain
was suggested to have an important role in weakening the even-filling correlated insulator states in MATBG~\cite{BultnickKhalaf2020,parker2020straininducedZalatel}.
This is mainly due to a large increase of the flat-bands bandwidth~\cite{TwistedGrapheneStrain,StrainDesignFu}, leading to diminished effective interactions.

\begin{figure}
\begin{centering}
\includegraphics[scale=0.45]{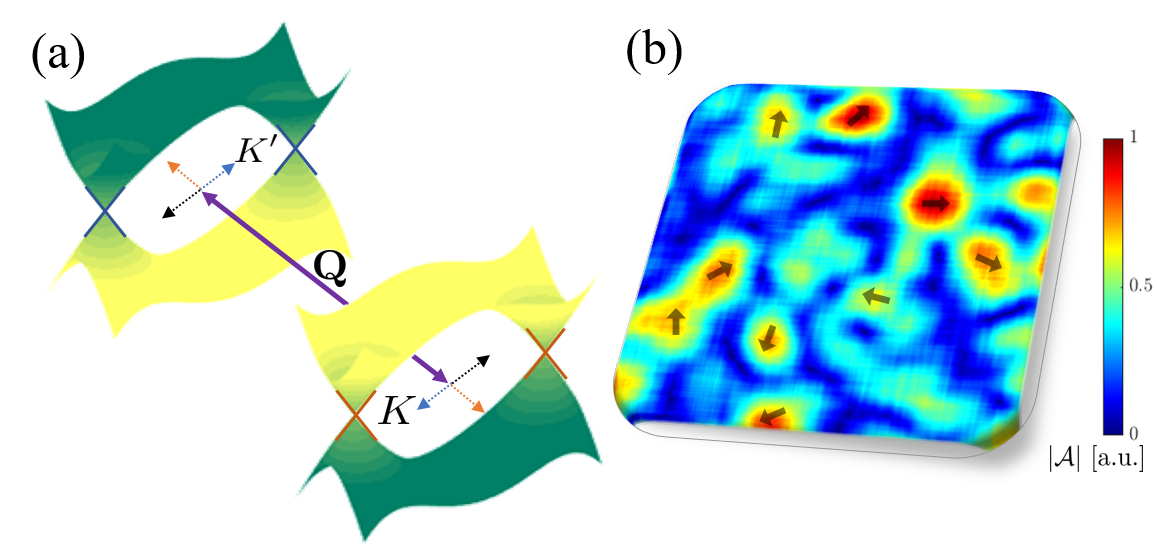}
\par\end{centering}
\caption{ \label{fig:schematicfigure}
Schematic description of our model.
(a) 
The band structure in each valley ($K,K'$) is approximated by two Dirac cones, one in each ``mini-valley".
The momentum separating the valleys, $\mathbf Q$, is modified by random homostrain fluctuations acting on the graphene layers, as represented by colored dotted arrows.
(b)
Schematic of a possible random strain configuration.
Color indicates the strain strength $\left|{\mathbfcal A}\right|$, whereas arrows indicate the directions of the local distortions of the graphene lattice.
}
\end{figure}

Whereas heterostrain shifts the two Dirac points within each valley with respect to each other, homostrain  subjects both layers to identical distortions and mainly acts as a pseudo-gauge field. This field acts oppositely in the two valleys (as illustrated in Fig.~\ref{fig:schematicfigure}a), thus maintaining time-reversal symmetry (TRS).
In this manuscript, we explore the effects of spatially random homostrain (Fig.~\ref{fig:schematicfigure}b) on the properties of the K-IVC phase.
We show that this  perturbation induces ``pair-breaking'' effects in striking resemblance to magnetic impurities in spin-singlet superconductors.
Effects of other types of strain on the K-IVC phase are discussed in the supplementary materials (SM)~\cite{SupplementRef}, and various disorder perturbations are classified by their impact on this phase in Ref.~\cite{AndersonTBG}. 

We find that modest strain fluctuations reduces both the K-IVC order parameter and the spectral gap, but enable the two to be dramatically different from one another. We propose that this effect may be responsible for the surprisingly small activation gap observed in transport experiments.
Moreover, we show that intervalley coherence can persist even when the system becomes gapless to single-particle excitations, a phase that we dub  ``gapless K-IVC''.
This phase could explain the haphazard appearance of a correlated insulator at CN.

\textit{Model.---}We consider the following simplified spinless model of MATBG,
\begin{equation}
    H_0=\sum_{\mathbf k} 
    \Psi^\dagger_ {\mathbf k}
    \left[
    u k_x \sigma_x \tau_z + u k_y\sigma_y
    \right]
    \Psi^{\phantom{\dagger}}_{\mathbf k},\label{eq:H0model}
\end{equation}
where $\Psi^\dagger_ {\mathbf k}$ is an 8-spinor of fermionic annihilation operators at momentum $\mathbf k$ relative to their respective origin in momentum space, and with sublattice, valley, and "mini-valley" degrees of freedom, described by Pauli matrices $\sigma_i$, $\tau_i$, and $\rho_i$, respectively.
This model features four Dirac cones with linear dispersion $\epsilon_{\mathbf k}=\pm u \left|{\mathbf k}\right|$.
$H_0$ preserves the TRS ${\cal T}=\tau_x\rho_x{\cal K}$ (${\cal K}$ implements
complex conjugation), with ${\cal T}^2=1$.

The spinless model effectively describes the fillings $\nu=\pm2$, where two electronic flavors of opposite valleys are entirely filled or empty, while the remaining pair is ``active'' and forms intervalley coherence.
Conversely, two copies of our spinless model may be used describe the K-IVC phase at CN~\cite{BultnickKhalaf2020,hofmann2021fermionic,ShavitMaATBGprl,OxfordMATBG}.

We introduce local density-density repulsive interactions,
\begin{equation}
    H_{\rm int}=\frac{U}{2\Omega}
    \sum_{{\mathbf k},{\mathbf k'},{\mathbf q}}
    \Psi^\dagger_ {\mathbf {k+q}}\Psi^{\phantom{\dagger}}_ {\mathbf k}
    \Psi^\dagger_ {\mathbf {k'-q}}\Psi^{\phantom{\dagger}}_ {\mathbf {k'}}
    \label{eq:Hintterm},
\end{equation}
which induce spontaneous symmetry-breaking in our model on a mean-field level. ($\Omega$ is the system area.) This simplified form of $H_{\rm int}$ suffices to illustrate the phenomenon we are interested in.
Specifically, we examine the K-IVC phase, which has been argued to be the likely ground state of MATBG at even fillings. 
The K-IVC state is characterized by intervalley coherence, i.e.,  formation of an exciton condensate with intervalley electron-hole pairs, a finite gap to charge excitations, and TRS breaking. 
The corresponding mean-field Hamiltonian at a given $\mathbf k$ has the form
\begin{equation}
    h_{\rm MF}\left({\mathbf k}\right) = u\left(k_x\sigma_x\tau_z+k_y\sigma_y\right)+\Delta_{\rm ivc} \sigma_x\tau_x\rho_z.\label{eq:hMFmeanfield}
\end{equation}
$ h_{\rm MF}$ preserves a Kramers-like TRS, ${\cal T}'=\tau_y\rho_x{\cal K}$, which concatenates $\mathcal{T}$ with a valley rotation.
It also preserves chiral and particle-hole symmetries, represented by ${\cal S}=\sigma_z$ and ${\cal C}={\cal S}{\cal T}'$, respectively~
\footnote{We note that a more realistic model of the band structure of MATBG contains terms which break particle-hole symmetry, as well as terms proportional to $\tau_z$. These break $\cal S$ and $\cal C$, reducing the symmetry of $h_{\rm MF}$ from symmetry class CII to AII.}~.

We now introduce random homostrain variations, which enter the model as a random gauge field acting in opposite directions in opposite valleys, see Fig.~\ref{fig:schematicfigure}b. Concretely, the strain Hamiltonian may be written as
\begin{equation}
    H_{\rm str}=u\int d{\mathbf r}
    \Psi^\dagger\left({\mathbf r}\right)
    \left[
    {\cal A}_x \left({\mathbf r}\right)\sigma_x + {\cal A}_y \left({\mathbf r}\right)\sigma_y\tau_z
    \right]
    \Psi\left({\mathbf r}\right),\label{eq:StrainHamiltonianPerturbation}
\end{equation}
where $\mathbfcal{A}$
is the strain-induced perturbation, and $\Psi\left({\mathbf r}\right)=\frac{1}{\sqrt{\Omega}}\sum_{\mathbf k}e^{i{\mathbf k}\cdot{\mathbf r}} \Psi_{\mathbf k}$.
In Eq.\ \eqref{eq:StrainHamiltonianPerturbation} we have implicitly assumed that the strain potential is smooth on the moir\'e length scale. Otherwise one should replace the renormalized velocity $u$ by the much larger bare graphene Fermi velocity $v_F$.
Notice that uniform strain constitutes a shift of $\mathbf k$ in $H_0$~\cite{gaugefieldsingraphenepaco} and redefines the momentum connecting the two valleys $\mathbf{Q=K-K'}$. This only changes the momentum carried by the condensed electron-hole K-IVC pairs, so that we can assume ${\cal A}_i$ has zero spatial mean. 

The perturbation in Eq.~\eqref{eq:StrainHamiltonianPerturbation} does not break any of the symmetries of $h_{\rm MF}$.
However, as we will show below (see also Ref.\ \cite{AndersonTBG}), the fact that it commutes with the K-IVC operator $\sigma_x\tau_x\rho_z$, enables a drastic reduction of the gap which opens in the K-IVC spectrum due to the random strain disorder. 
For this reason, we have also neglected inter-minivalley scattering, which have additional $\rho_x,\rho_y$ factors, and thus anticommute with the order parameter.

Considering a simple point-like perturbation, $\mathbfcal{A}\left({\mathbf r}\right)=\mathbfcal{A}_0 \delta\left({\mathbf r}\right)$, one may employ a $T$-matrix formalism to find the bound-state spectrum inside the mean-field gap~\cite{SupplementRef}, in analogy to Yu-Shiba-Rusinov states induced by magnetic impurities in a superconductor~\cite{Yu1965,Shiba1968,Rusinov1969}. We find that as the perturbation strength increases, the in-gap-state energy is reduced and the two bound-state energies cross at zero when the impurity strength becomes an appreciable fraction of the bandwidth $W$.

Moreover, we find that when approximating the density-of-states (DOS) as a constant around the Fermi level (rather than linear as appropriate in our case), one recovers -- apart from additional degeneracies -- precisely the bound-state spectrum of a magnetic impurity inside a singlet superconductor.
This outcome may be traced to the analogous algebraic structure of the two problems, i.e., the impurity operator commuting with the order parameter.
This analogy enables the treatment of random strain fluctuations in MATBG by tools similar to those employed in superconductors with magnetic impurities.

\textit{Abrikosov-Gor'kov approach.---}We therefore treat the random strain fluctuations within the self-consistent Born approximation (SCBA), inspired by the Abrikosov-Gor'kov theory of superconductivity in magnetically disordered alloys~\cite{AGmethodAbrikosov1959theory}. A similar method was also used to study exciton condensates in the presence of potential impurities~\cite{AGmethodZittartz,AGmethodBistritzerMacdonald}.

The main object of interest is the Green's function,
\begin{equation}
    G\left({\mathbf k},\omega\right) = \left(i\omega-h_{\rm MF}- \hat{\Sigma}_{\rm SCBA}\right)^{-1}.\label{eq:GreensFunction}
\end{equation}
Within the SCBA, the self-energy matrix $ \hat{\Sigma}_{\rm SCBA}$ can be written as
\begin{equation}
     \hat{\Sigma}_{\rm SCBA}\left({\mathbf k},\omega\right)=
    \left\langle
    \sum_{\mathbf p}
    {\cal U}_{{\mathbf k}-{\mathbf p}}
    G\left({\mathbf p},\omega\right)
    {\cal U}_{{\mathbf p}-{\mathbf k}}
    \right\rangle_{\rm dis.},
    \label{eq:SCBAmatrix} 
\end{equation}
where the matrix $\cal U$ represents the random strain perturbation in momentum space, $H_{\rm str} = \sum_{\mathbf{kq}}\Psi^{\dagger}_{\mathbf{k+q}}{\cal U}_\mathbf{q}\Psi_\mathbf{k}$,
and $\left\langle...\right\rangle_{\rm dis.}$ stands for disorder averaging.

Upon standard manipulation of the Green's function, it can be written as
\begin{equation}
    G=-\frac{i\tilde{\omega}+u\left(k_x\sigma_x\tau_z+k_y\sigma_y\right)+\tilde{\Delta} \sigma_x\tau_x\rho_z}{\epsilon_{\mathbf k}^{2}+\tilde{\Delta}^{2}+\tilde{\omega}^{2}}.\label{eq:explicitGreensfunction}
\end{equation}
The parameters $\tilde{\omega},\tilde{\Delta}$ are related to ${\omega},{\Delta_{\rm ivc}}$ by the self-consistency equations
\begin{equation}
    \begin{pmatrix}\tilde{\omega}\\
\tilde{\Delta}
\end{pmatrix}=\begin{pmatrix}\omega\\
\Delta_{{\rm ivc}}
\end{pmatrix}+\Gamma{\cal F}\left(\tilde{\omega},\tilde{\Delta}\right)\begin{pmatrix}\tilde{\omega}\\
-\tilde{\Delta}
\end{pmatrix},
     \label{eq:tildesEquation}
\end{equation}
where ${\cal F}\left(\tilde{\omega},\tilde{\Delta}\right)=\frac{W}{2}\int d\epsilon\frac{{\cal N}\left(\epsilon\right)}{\epsilon^2+\tilde{\Delta}^2+\tilde{\omega}^2}$, ${\cal N}\left(\epsilon\right)$ is the DOS per unit cell of area $\Omega_{u.c.}$, and $W$ is the bandwidth. The disorder energy scale $\Gamma$ is
\begin{equation}
    \Gamma = \frac{2\Omega}{W \Omega_{u.c.}}\left\langle \int d\theta  u^2 \mathbf{A}_\theta^\dagger \cdot\mathbf{A}_\theta^{\phantom{\dagger}}
    \right\rangle_{\rm dis.}\label{eq:Gammamain},
\end{equation}
where $\mathbf{A}_\mathbf{q}$ is the Fourier transform of $\mathbfcal{A}\left(\mathbf{r}\right)$, and we have used the standard approximation that the scattering mostly depends on the angle between incoming and outgoing momenta $\theta$~\cite{AGmethodZittartz,AGmethodBistritzerMacdonald}. As for magnetic impurities in superconductors and potential scatterers in excitonic condensates, the form of Eq.~\eqref{eq:tildesEquation} is due to the K-IVC order parameter commuting with the random perturbation.
Thus, the equations we find are identical to the Abrikosov-Gor'kov equations for superconductors with magnetic impurities, with one important difference. In MATBG, we cannot assume a constant DOS near the Fermi energy, but should account for the fact that the DOS \textit{vanishes linearly} at the Dirac point. This leads to important qualitative differences in the results.

By relating the local strain to the effective gauge-field $\mathbfcal{A}$, one may obtain an order-of-magnitude estimate of $\Gamma$.
For root-mean-square strains of ${\cal E} \sim 0.1\%$ and disorder correlation lengths of few unit cells, we find $\Gamma/W$ values of $0.1-0.3$~\cite{SupplementRef}.
As will be shown, such values are sufficient to dramatically reduce the spectral gap or even close it completely.

\begin{figure}
\begin{centering}
\includegraphics[scale=0.52]{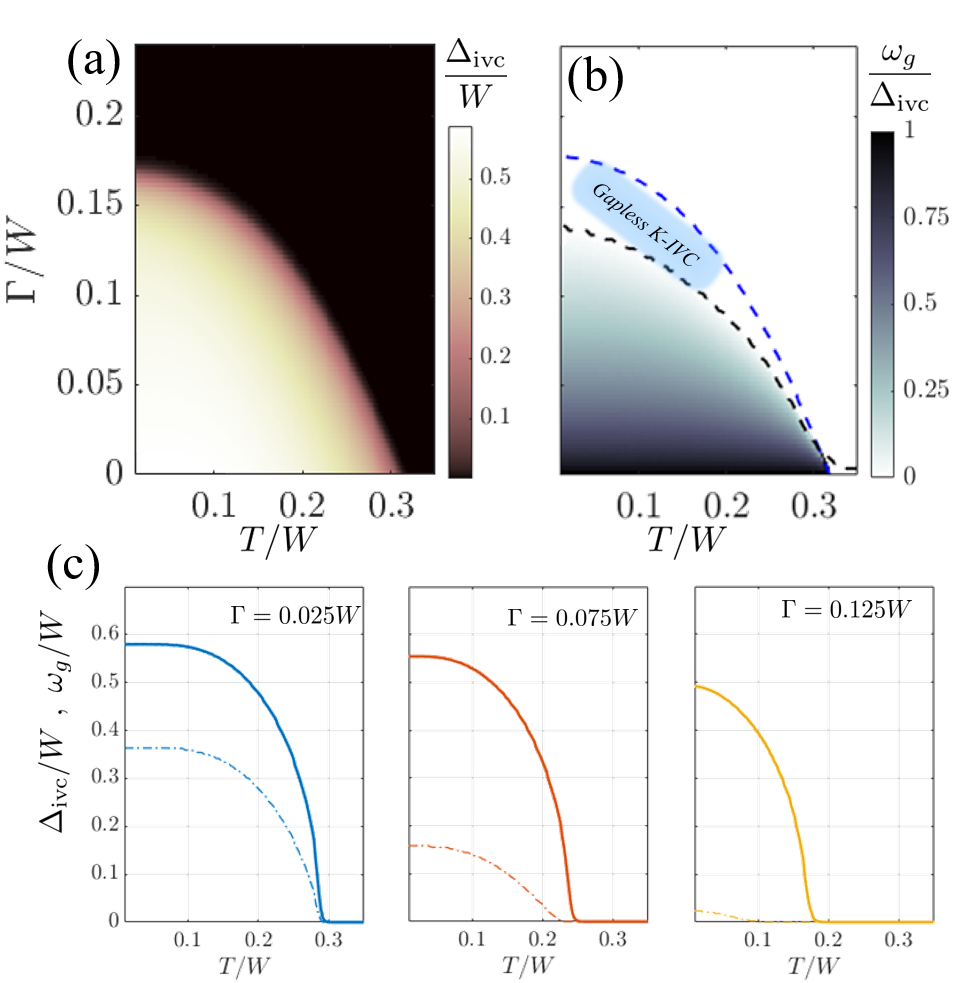}
\par\end{centering}
\caption{ \label{fig:DeltaOmegagMaps}
(a) The order parameter $\Delta_{\rm ivc}$ as a function of temperature and disorder energy scale $\Gamma$.
For a given temperature, there is a critical $\Gamma$ at which the order parameter vanishes.
(b) The ratio between the gap in the TDOS, $\omega_g$, and  $\Delta_{\rm ivc}$.
As $\Gamma$ increases, $\omega_g$ deviates from $\Delta_{\rm ivc}$ appreciably.
There exists a gapless K-IVC region bounded by the vanishing of the single-particle gap (black dashed line) and of the order parameter (blue dashed line).
We used $U=0.7W$ and a linear DOS ${\cal N}_{\rm lin.}$.
(c) Linecuts for different values of $\Gamma$ showing $\Delta_{\rm ivc}$ (solid) and $\omega_g$ (dash-dotted). 
}
\end{figure}

The strength of K-IVC order in the presence of disorder is obtained by combining Eq.~\eqref{eq:tildesEquation} with the gap equation
\begin{equation}
    \Delta_{\rm ivc}=-2\frac{U}{\beta\Omega}
    \sum_{\omega{\mathbf k}}{\rm Tr}\left\{ \sigma_{x}\tau_{x}\rho_{z}G\right\},\label{eq:selfconsistentgapequation}
\end{equation}
which we solve numerically. (Here, $\beta$ is inverse temperature.) Figure \ref{fig:DeltaOmegagMaps}a shows results for the order parameter $\Delta_{\rm ivc}$ as a function of temperature and disorder. We find that both the order parameter and the critical temperature deteriorate with increasing disorder.

When assuming a DOS which is linear in energy, ${\cal N}_{\rm lin.}=4\left|\epsilon\right|/W^2$ with cutoff energy  $\left|\epsilon\right|<W/2$, we can also make analytical progress. 
In particular, we may calculate the critical disorder scale $\Gamma_c$, at which $\Delta_{\rm ivc}$ vanishes for $T=0$. In the regime $\Gamma_c\ll W/2$, we find  ~\cite{SupplementRef}
\begin{equation}
    \Gamma_c=\frac{U_{c}}{\log8}\left(1-\frac{U_{c}}{U}\right),\label{eq:criticalGammac}
\end{equation}
where $U_c=W/4$ is the critical interaction $U$ below which the K-IVC order vanishes at $\Gamma,T=0$.
The finite $U_c$ as well as the form of the critical ``pair-breaking'' parameter $\Gamma_c$ originate in the DOS vanishing linearly at zero. This suppresses the analog of the Cooper-instability for arbitrarily weak interactions, which requires a finite DOS at the Fermi level.

Having found the self-consistent Green's function, we can calculate the tunneling density of states (TDOS)  $\tilde{\cal N} \left(\epsilon\right) = \frac{1}{\pi}{\rm Im} \frac{1}{\Omega}\sum_{\mathbf k} {\rm Tr} G\left({\mathbf k},\omega\to i\epsilon\right)$.
In clean systems, the gap $\omega_g$ in the TDOS is equal to the order parameter $\Delta_{\rm ivc}$. 
In the presence of finite disorder $\Gamma$, $\omega_g$ is in general smaller than $\Delta_{\rm ivc}$. 

In Fig.~\ref{fig:DeltaOmegagMaps}b, we plot the ratio $\omega_g/\Delta_{\rm ivc}$ for the same parameter range as in Fig.~\ref{fig:DeltaOmegagMaps}a. As $\Gamma$ increases, the ratio gradually deviates from unity, and eventually reaches zero, \textit{before} $\Delta_{\rm ivc}$ vanishes. We dub the regime with finite $\Delta_\mathrm{ivc}$ and vanishing $\omega_g$ as the gapless K-IVC phase.
Similar to gapless superconductivity, we interpret this regime as one where intervalley coherence exists throughout a large fraction of the system, yet strain-induced in-gap states form a low-energy compressible continuum of states.
For linear DOS, we find that the disorder strength at which the gap closes is related to $\Delta_{\rm ivc}$~\cite{SupplementRef} through $\Gamma_g=W/\log\left(1+\frac{W^{2}}{\Delta_{{\rm ivc}}^{2}}\right)^2$.

In Fig.~\ref{fig:DeltaOmegagMaps}c, we plot linecuts of Fig.~\ref{fig:DeltaOmegagMaps}a for several values of~$\Gamma$. We find that even for modest (and realistic) disorder strengths, the spectral gap is significantly suppressed compared to the naive gap as given by the order parameter.
Thus, devices which host appreciable strain fluctuations
may exhibit an effective gap as seen in a global transport measurement, which is about an order of magnitude smaller than the expected condensation order parameter.

\begin{figure}
\begin{centering}
\includegraphics[scale=0.57]{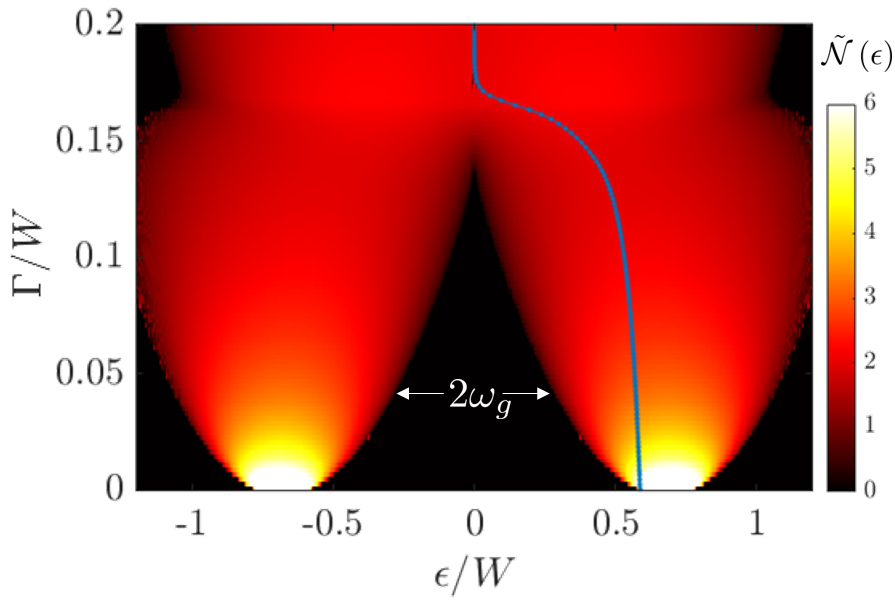}
\par\end{centering}
\caption{ \label{fig:DOSmap}
Evolution of the TDOS $\tilde{\cal N}(\epsilon)$ with $\Gamma$.
The gap $\omega_g$ gradually closes as the disorder strength increases.
The blue line indicates the evolution of $\Delta_{\rm ivc}/W$ with increasing $\Gamma$. 
Notice that its evolution differs from that of the single-particle gap as it deteriorates more slowly.
We used $U=0.7W$, $T=0$ in this figure.
}
\end{figure}

In Fig.~\ref{fig:DOSmap} we track the evolution of the TDOS with disorder strength.
At very low $\Gamma$ we find two narrow K-IVC bands separated by roughly $2\Delta_{\rm ivc}$, as expected from previous theoretical investigations of the pristine K-IVC phase. As the disorder strength $\Gamma$ increases, these bands spread out in energy, and their separation diminishes. As shown in the figure, this is very different from the behavior of $\Delta_{\rm ivc}$. While~$\omega_g$ diminishes already for weak disorder, 
$\Delta_{\rm ivc}$ remains mostly unaffected up to intermediate values of $\Gamma$.

The TDOS depicted in Fig.~\ref{fig:DOSmap} can be measured in planar tunneling junctions with a large tunneling area, similar to the experimental verification of gapless superconductivity~\cite{PlanarTDOSReif}.
The large tunneling area is required for effective averaging over disorder configurations in a particular device.
Such measurements are expected to show two spread-out TDOS lobes, with centers separated by $\sim 2\Delta_{\rm ivc}$ and a spectral gap of $2\omega_g<2\Delta_{\rm ivc}$. 
In contrast, local scanning-tunneling-microscopy (STM) measurements do not reveal disorder-averaged quantities, yet we expect the tunneling gap to vary as a function of position, in a manner correlated with the homostrain variations. Such phenomenology would be a clear indicator of the proposed disordered K-IVC phase.

\begin{figure}
\begin{centering}
\includegraphics[scale=0.7]{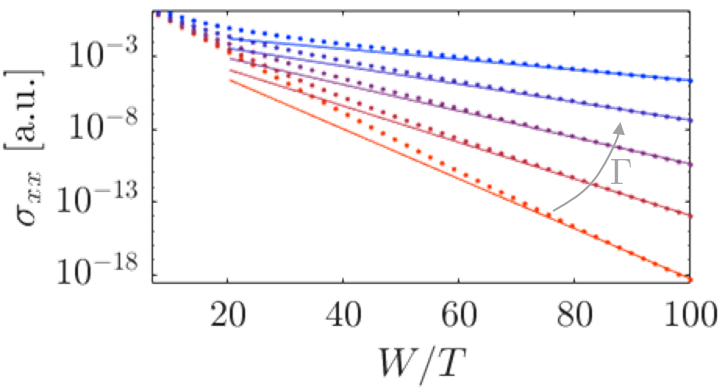}
\par\end{centering}
\caption{ \label{fig:activationTransport}
Arrhenius plots of the longitudinal DC conductivity $\sigma_{xx}$. 
Dots: $\sigma_{xx}$ as a function of temperature for different strengths of the strain fluctuations, $\Gamma/W\in\left[0.02,0.1\right]$, increasing from bottom (red) to top (blue) in steps of $0.02$. We used the same parameters as in Fig.~\ref{fig:DeltaOmegagMaps}.
Solid lines: guides to the eye $\propto\exp\left[ -\omega_g\left(T\to 0\right)/T\right]$ for each $\Gamma$ value. Whereas $\omega_g$ differs by a factor of $\sim 5$ between the first and last plots, $\Delta_{\rm ivc}$ changes only modestly by $< 25\%$.
}
\end{figure}

To make contact with transport experiments, we now turn to calculating the DC conductivity within our model. We use the Kubo-Bastin  formula~\cite{KuboBASTIN19711811}, $\sigma_{xx}\propto\int d\epsilon\left(-\frac{df}{d\epsilon}\right){\cal S}\left(\epsilon\right)$, where $f$ is the Fermi function, and 
\begin{equation}
     {\cal S}\left(\epsilon\right)
    \approx
    \frac{1}{\Omega}\sum_{\mathbf{k}}
    {\rm Tr}\left\{
    j_x
    {\rm Im}G\left({\mathbf k},\omega\to i\epsilon\right)
    j_x
    {\rm Im}G\left({\mathbf k},\omega\to i\epsilon\right)
    \right\} \label{eq:Sepsilonnovertexcorr}
\end{equation}
is the conductivity kernel. The current operator is $j_x=u\sigma_x\tau_z$. Equation~\eqref{eq:Sepsilonnovertexcorr} neglects vertex corrections to $j_x$, which may have quantitative significance, but are beyond the scope of this work.

Fig.~\ref{fig:activationTransport} presents Arrhenius plots of the conductivity for different $\Gamma$ values. From these plots, we can extract the activation energy 
$E_{\rm act}$ by fitting the longitudinal conductivity to $\sigma_{xx}\propto\exp\left(-E_{\rm act}/T\right)$.
Remarkably, we observe that the low-$T$ behavior is indeed temperature activated with $E_{\rm act}\approx \omega_g$, and not the potentially much larger $\Delta_{\rm ivc}$. (In the topmost plot of Fig.~\ref{fig:activationTransport} $\Delta_{\rm ivc}\approx 5.4 \omega_g$.) 
This behavior can be traced to the analytic structure of ${\cal S}\left(\epsilon\right)$, which has a gap $\approx 2\omega_g$ around $\epsilon=0$~\cite{SupplementRef}, similar to the TDOS.

\textit{Conclusions.---}We have explored the consequences of random homostrain fluctuations on the K-IVC state, believed to describe the insulating phases of MATBG at even fillings.
Using a simplified model for MATBG,
we have studied this problem using the SCBA in conjunction with a mean-field treatment of the K-IVC order parameter.
Homostrain disorder has a pair-breaking effect on the intervalley coherent condensate, since it locally acts on the two valleys in opposite ways.

In contrast to similar pair-breaking disorder problems, random homostrain does not break any symmetries of the K-IVC state.
However, it does lead to in-gap states, gap closing, and order parameter deterioration due to its operator structure -- it commutes with the order parameter.
Moreover, the DOS dependence on energy had to be taken into account, since it vanishes at the Dirac point.
This led to the unique form of the solutions of the Abrikosov-Gor'kov equations which we derive, and of the critical pair-breaking parameter $\Gamma_c$.

One of our key results is the significant separation between the energy scales of the K-IVC order ($\Delta_\mathrm{ivc}$) and the spectral gap for single-particle excitations ($\omega_g$), even for modest values of disorder.
Borrowing insights from superconductors, the gap reduction stems from in-gap bound states, which become stronger and more abundant with increasing~$\Gamma$, yet impact the surrounding intervalley-coherent condensate only weakly.

We suggest that the order-of-magnitude discrepancy between the theorized K-IVC gap and the activation gap measured in transport experiments can be resolved within our model. 
We have demonstrated that the relevant activation energy as measured via the DC conductivity is the spectral gap, which may be considerably smaller than the order parameter due to disorder. (Both scales coincide in the pristine case).
The rare appearance of insulators at CN can also be understood by considering two copies of our model with different spin labels. 
Variations of the magnitude of strain disorder between devices may tip the state at CN from a weakly-insulating K-IVC state to the gapless K-IVC regime.
The relative weakness of the insulating state at CN compared to fillings $\nu=\pm 2$ has been attributed to bandwidth renormalizations~\cite{BandStructurenoramlizations,HartreeBandStructure,Mcdonaldweakfieldhall}, rendering the effective interactions stronger away from CN.

The interplay of strain fluctuations with other sources of K-IVC suppression, such as twist-angle disorder and uniform heterostrain, remains to be explored.
Additionally, the fact that the considered disorder couples only to intervalley-ordered states may also be important.
This may have ramifications for the competition between the K-IVC state and other correlated insulating phases, such as the valley-Hall phase, for which the order parameter  $\propto\sigma_z$  anticommutes with the strain fluctuations.
Incorporating such complications, as well as including more intricate aspects of the (particle-hole asymmetric) band structure will shed much-needed light on the nature of the insulating phases in MATBG and their variation between different devices.

\begin{acknowledgments}
We gratefully acknowledge funding by Deutsche Forschungsgemeinschaft through CRC 183 (project C02; YO and FvO), a joint ANR-DFG project  (TWISTGRAPH; CM and FvO), the European Union’s Horizon 2020 research and innovation program (Grant Agreement LEGOTOP No. 788715; YO), the ISF Quantum Science and Technology program (2074/19; YO), and a BSF and NSF grant (2018643; YO).
\end{acknowledgments}

\bibliographystyle{apsrev4-2}
\bibliography{StrainFluctIVCref.bib}

\begin{widetext}
\setcounter{section}{0} \renewcommand{\thesection}{S.\arabic{section}} \setcounter{figure}{0} \renewcommand{\thefigure}{S\arabic{figure}} \setcounter{equation}{0} \renewcommand{\theequation}{S\arabic{equation}}

\section* {Supplemental material}

\section{Components of the strain tensor}
Consider the two graphene layers which make up the twisted bilayer.
One may apply a spatially-dependent displacement field to the carbon atoms of each layer, $u^\alpha_{\rm displacement}\left(\ell,{\mathbf r}\right)$.
Here, $\mathbf r$ is the position of the displaced carbon atom, $\ell={\rm t},{\rm b}$ corresponds to either the top or bottom layer, and $\alpha=x,y$ is the axis along which the atom is displaced.

We then define the strain tensor
\begin{equation}
    \epsilon^{\ell}_{\alpha\beta}=\frac{\partial_\alpha u_{\rm displacement}^\beta\left(\ell,{\mathbf r}\right) + \partial_\beta u_{\rm displacement}^\alpha\left(\ell,{\mathbf r}\right)}{2}.\label{eq:suppstraintensorbylayer}
\end{equation}
The homostrain and heterostrain tensor are now simply the layer-symmetric and antisymmetric components of $\epsilon^{\ell}_{\alpha\beta}$,
\begin{equation}
    \epsilon^{\rm homo}_{\alpha\beta} =\frac{ \epsilon^{\rm t}_{\alpha\beta}+\epsilon^{\rm b}_{\alpha\beta}}{2},\label{eq:epsilonhomosupp}
\end{equation}
\begin{equation}
    \epsilon^{\rm hetero}_{\alpha\beta} =\frac{ \epsilon^{\rm t}_{\alpha\beta}-\epsilon^{\rm b}_{\alpha\beta}}{2},\label{eq:epsilonheterosupp}
\end{equation}
and one therefore expects both components to exist for a generic random strain field configuration. 

\subsection{Uniform strain}

\textit{Uniform heterostrain.---}The effects of uniform heterostrain on both the band structure of magic angle twisted bilayer graphene (MATBG) and on the correlated insulator appearing at even fillings of electron/holes per moir\'e unit-cell have been considered in Refs.~\cite{BultnickKhalaf2020,parker2020straininducedZalatel,TwistedGrapheneStrain,StrainDesignFu}.
The main notable effect of this perturbation is the distortion of the moir\'e superlattice, leading to a significant increase of the flat-bands' bandwidth. This then leads to weakening of the correlated insulator, due to an effective weaker interaction (as compared to the bandwidth of the electrons).

\textit{Uniform homostrain.---}Uniform strain which is applied identically to both graphene layers will shift the band structure in the two valleys in opposite directions in momentum space~\cite{gaugefieldsingraphenepaco},
\begin{equation}
    {\mathbf K} \to {\mathbf K} + {\mathbf A}, \,\,\,\,
    {\mathbf K'} \to {\mathbf K'} - {\mathbf A},
\end{equation}
where ${\mathbf K},{\mathbf K'}$ are the two monolayer graphene valleys and ${\mathbf A}$ is the pseudo-gauge field.

Assuming electron-electron interactions which are weakly dependent on momentum, consider a spontaneous intervalley-coherent order parameter, which has an oscillatory spatial dependence $\Delta_{\rm ivc}\left({\mathbf r}\right)\propto \exp\left(i{\mathbf Q}\cdot {\mathbf r}\right)$ in the absence of strain, with ${\mathbf Q}={\mathbf K}-{\mathbf K'}$ the intervalley separation.
Then, applying uniform homostrain would have no effect on the instability towards the intervalley order, though the ``new'' order parameter will now have a slightly different spatial periodicity, $\Delta_{\rm ivc}\left({\mathbf r}\right)\propto \exp\left[i\left({\mathbf Q+2{\mathbf A}}\right)\cdot {\mathbf r}\right]$.
We therefore do not include this type of perturbation in our considerations in this work.

\subsection{Strain disorder}
Let us now discuss the effects of the disordered strain potentials within our effective model.
As a reminder, its mean-field form is
\begin{equation}
     h_{\rm MF}\left({\mathbf k}\right) = u\left(k_x\sigma_x\tau_z+k_y\sigma_y\right)+\Delta_{\rm ivc} \sigma_x\tau_x\rho_z,\label{eq:meanfieldsupp}
\end{equation}
where sublattice, valley, and "mini-valley" degrees of freedom are described by Pauli matrices $\sigma_i$, $\tau_i$, and $\rho_i$, respectively.

In this language, homostrain disorder is proportional to the operators $\sigma_x$, $\sigma_y\tau_z$, or a linear combination of the two.
Therefore, the homostrain operator commutes with the order parameter $\propto \sigma_x\tau_x\rho_z$.
This fact is crucial for allowing this disorder term to close the gap induced by the intervalley order parameter, as we discuss throughout this work.

Heterostrain disorder however, is proportional to the operators $\sigma_x\tau_z\rho_z$, $\sigma_y\rho_z$ and their combinations.
One easy way to understand why, is to recognize that the same mini-valley in opposite valleys originates in the Dirac points corresponding to opposite layers.
Therefore, when it comes to response to a strain potential, the operator $\tau_z\rho_z$ effectively acts as an ``originating-layer'' index.
Heterostrain operators are then naturally given by multiplying the homostrain operators by $\tau_z\rho_z$, enforcing opposite pseudo-gauge fields in opposite layers.

Importantly, the heterostrain operators anticommute with the intervalley-coherent order parameter, so that the gap cannot be closed by weak disorder of that kind.
Their effect is thus analogous to that of non-magnetic charge impurities in conventional spin-singlet superconductors.
Therefore, we do not consider this type of disorder in this work.

\section{Mean-field results in the pristine case}
Let us briefly discuss the model
\begin{equation}
    H=H_0+H_{\rm int},
\end{equation}
\begin{equation}
    H_0=\sum_{\mathbf k} 
    \Psi^\dagger_ {\mathbf k}
    \left[
    u k_x \sigma_x \tau_z + u k_y\sigma_y
    \right]
    \Psi _{\mathbf k},\label{eq:H0modelsupplemantary}
\end{equation}
\begin{equation}
    H_{\rm int}=\frac{U}{2\Omega}
    \sum_{{\mathbf k},{\mathbf k'},{\mathbf q}}
    \Psi^\dagger_ {\mathbf {k+q}}\Psi_ {\mathbf k}
    \Psi^\dagger_ {\mathbf {k'-q}}\Psi_ {\mathbf {k'}}
    \label{eq:Hinttermsupplemantary}.
\end{equation}
$\Psi^\dagger_ {\mathbf k}$ is an 8-spinor of fermionic annihilation operators at momentum $\mathbf k$ relative to their respective origin in momentum space, and with sublattice, valley, and "mini-valley" degrees of freedom, described by Pauli matrices $\sigma_i$, $\tau_i$, and $\rho_i$, respectively.

We note that the basis used in this model is quite different from the commonly used ``sublattice basis'' (see, e.g., Refs. \cite{BultnickKhalaf2020,ShavitMaATBGprl}), since we effectively separate the electronic species in each valley into two species near the two Dirac cones in each valley.
In this case, the momentum $\mathbf{k}$ is measured relative to the appropriate Dirac point labeled by both the quantum numbers $\tau_z$ and $\rho_z$.
As we show momentarily, this also translates to a different form of the intervalley-coherent order parameter.

Our model is meant to provide an effective description of the physics near $\nu=\pm 2$ filling in MATBG.
It is widely believed that in these regimes, the flavor (spin and valley) symmetry is spontaneously broken due to strong electron-electron interactions~\cite{DiracRevivals,YazdaniRevivals}.
Consequently, two flavors are either completely full or completely empty, whereas the remaining two flavors have their chemical potential at the Dirac point.
Eq.~\eqref{eq:H0modelsupplemantary} describes precisely the two remaining active flavors in this scenario.

We are interested in the order parameter  
\begin{equation}
    \Delta_{{\rm ivc}}=-\frac{U}{2\Omega}\sum_{\mathbf{k}}\left\langle \Psi^{\dagger}\left(\mathbf{k}\right)\sigma_{x}\tau_{x}\rho_{z}\Psi\left(\mathbf{k}\right)\right\rangle
\end{equation}
for Kramers intervalley-coherent (K-IVC) order.
We note that $\Delta_{{\rm ivc}}$ spontaneously breaks the valley-$U\left(1\right)$ symmetry in our model, manifested in $\tau_x\to \tau_x e^{i\alpha_{\rm valley}\tau_z}$.

Performing a mean-field decomposition on the Hamiltonian, we find
\begin{equation}
    H_{{\rm MF}}=\sum_{\mathbf{k}}\Psi^{\dagger}\left(\mathbf{k}\right)\left[uk_{x}\sigma_{x}\tau_{z}+uk_{y}\sigma_{y}+\Delta_{{\rm ivc}}\sigma_{x}\tau_{x}\rho_{z}\right]\Psi\left(\mathbf{k}\right)+2\frac{\Omega}{U}\Delta_{{\rm ivc}}^{2}.\label{eq:meanfielddecompositionsupplementary}
\end{equation}
Importantly, the order parameter $\Delta_{\rm ivc}$ is the only possible K-IVC order parameter (up to a rotation in the $\tau_x$--$\tau_y$ plane), i.e., it fulfills the three conditions: (i) it is valley off-diagonal, (ii) it opens a gap in the spectrum (anti-commutes with $H_0$), and (iii) it breaks the time-reversal symmetry ${\cal T}=\tau_x\rho_x{\cal K}$, while preserving the Kramers one ${\cal T}'=\tau_y\rho_x{\cal K}$ ($\cal K$ is the complex conjugation operator).

Defining $\epsilon_\mathbf{k}=u\left|{\mathbf k}\right|$, $E_\mathbf{k}=\sqrt{\epsilon_\mathbf{k}^2+\Delta_{\rm ivc}^2}$, we find the mean-field thermodynamic potential $\Phi$ at temperature $T$,
\begin{equation}
    \Phi=2\frac{\Omega}{U}\Delta_{{\rm ivc}}^{2}-8T\sum_{\mathbf{k}}\log\left[2\cosh\left(\frac{E_{\mathbf{k}}}{2T}\right)\right].\label{eq:thermodynamicpotentialsupplementary}
\end{equation}
The self-consistent gap equation follows from the condition $\frac{\partial\Phi}{\partial\Delta_{{\rm ivc}}}=0$, which leads to
\begin{equation}
    \frac{1}{U}=\int d\epsilon{\cal N}\left(\epsilon\right)\frac{\tanh\left(\frac{\sqrt{\epsilon^{2}+\Delta_{{\rm ivc}}^{2}}}{2T}\right)}{\sqrt{\epsilon^{2}+\Delta_{{\rm ivc}}^{2}}}\label{eq:selfconsistentpristinesupplementary},
\end{equation}
where ${\cal N}\left(\epsilon\right)$ is the density of states per unit cell and flavor (spin, valley, minivalley) of $H_0$ at energy $\epsilon$.

We may extract some analytical results by assuming a strictly linear density of states,
\begin{equation}
    {\cal N}\left(\epsilon\right)=\begin{cases}
\frac{4\left|\epsilon\right|}{W^{2}} & \left|\epsilon\right|<W/2,\\
0 & {\rm otherwise},
\end{cases}\label{eq:DOSsupplementary}
\end{equation}
with $W$ the total bandwidth. 
Setting $T=0$, we find 
\begin{equation}
    \Delta_{00}=\Delta_{\rm ivc}\left(T=0\right)=U\left[1-\left(\frac{U_{c}}{U}\right)^{2}\right].\label{eq:Delta00suplementary}
\end{equation}
Here, we have defined the critical interaction strength $U_{c}=W/4$, below which there is no gap opening.
The existence of a critical coupling contrasts with the familiar behavior of the standard Bardeen-Cooper-Schrieffer (BCS) gap equation, which has the same form as Eq.~\eqref{eq:selfconsistentpristinesupplementary}.
The difference originates from the density of states ${\cal N}\left(\epsilon\right)$. In the present case, the density of states vanishes at the Dirac point ($\epsilon=0$), which eliminates the  Cooper instability for infinitesimally weak interaction strengths.

We may also extract the critical temperature $T_{c0}$ by setting $\Delta_{\rm ivc}=0$ in the gap equation. This leads to the transcendental equation
\begin{equation}
    \frac{U_{c}}{U}=\frac{T_{c0}}{U_{c}}\log\left[\cosh\left(\frac{U_{c}}{T_{c0}}\right)\right],\label{eq:Tc0transcedentalsupplementary}
\end{equation}
which does not have a  $T_{c0}>0$ solution for $U<U_c$, as expected.
Close to the transition, when $T_{c0}\ll U_c$, we may approximate $T_{c0}\approx\frac{1}{\log2}\frac{U_{c}}{U}\left(U-U_{c}\right)$.

\section{Point-like perturbation}


In this section, we analyze the effect of a localized strain ``impurity'' on the mean-field K-IVC state.
We assume that the strain does not suppress the formation of intervalley coherent order for the sake of this discussion.
Thus, we investigate the Hamiltonian
\begin{align}
    H_{\rm imp}&=\sum_{\mathbf k} 
    \Psi^\dagger_ {\mathbf k}
    \left[
    u k_x \sigma_x \tau_z + u k_y\sigma_y + \Delta_{\rm ivc}\sigma_x\tau_x\rho_z
    \right]
    \Psi _{\mathbf k}\nonumber\\
    &+\frac{1}{\Omega}\sum_{{\mathbf k},{\mathbf k'}} 
    \Psi^\dagger_ {\mathbf k}
    \left[
    v_F {\cal A}_x \sigma_x + v_F {\cal A}_y\sigma_y\tau_z
    \right]
    \Psi _{\mathbf k'},\label{eq:Himpuritysupplemental}
\end{align}
where ${\cal A}_{x,y}$ are constants, reflecting the delta-potential like spatial dependence of ${\mathbfcal A}\left({\mathbf r}\right)$, to be determined by the strain tensor, see Eq.~\eqref{eq:gaugefromstrainsupplemental}.
Notice that the velocity in this impurity strain potential is the bare Fermi velocity of graphene $v_F$, and not the renormalized one $u$.
This is because the deformed potential is not smooth on the moir\'e scale, as we have assumed it is pointlike by definition. 
We note that much like in the case of the strain perturbation we consider in the main text, $H_{\rm imp}$ preserves both $\cal T$ and ${\cal T}'$ time-reversal symmetries.

The Green's function for the ${\cal A}_{x,y}=0$ system can be written as
\begin{equation}
    G^0\left({\mathbf k},\omega\right)=-\frac{\omega+u k_{x}\sigma_{x}\tau_{z}+u k_{y}\sigma_{y}+\Delta_{{\rm ivc}}\sigma_{x}\tau_{x}\rho_{z}}{\epsilon_{\mathbf{k}}^{2}+\Delta_{{\rm ivc}}^{2}-\omega^{2}},\label{eq:G0supplemental}
\end{equation}
with $\epsilon_\mathbf{k}=u\left|{\mathbf k}\right|$.
We also calculate
\begin{equation}
    g^0\left(\omega\right)\equiv\frac{1}{\Omega}\sum_{\mathbf k}G^0\left({\mathbf k},\omega\right)=
    -\left(\omega+\Delta_{{\rm ivc}}\sigma_{x}\tau_{x}\rho_{z}\right)
    C
    ,\label{eq:g0supplemental}
\end{equation}
\begin{equation}                    
    C=
    \int d\epsilon\frac{{\cal N}\left(\epsilon\right)}{\epsilon^{2}+\Delta_{{\rm ivc}}^{2}-\omega^{2}},\label{eq:Cintegrationsupplemental}
\end{equation}
which we will use momentarily.

\subsection{$T$-matrix formalism}
Our goal is to find the in-gap bound state spectrum due to the strain perturbation. In order to do so, we shall employ the $T$-matrix formalism~\cite{Tmatrixhirschfeld1986resonant,TmatrixRMPBalatsky}.
Within this formalism, in the case of a point-like perturbation at the origin, the real-space Green's function may be written as
\begin{equation}
    G\left({\mathbf r},{\mathbf r'};\omega\right)
    =
    G^0\left({\mathbf r},{\mathbf r'};\omega\right)
    +
    G^0\left({\mathbf r},{\mathbf 0};\omega\right)
    T\left(\omega\right)
    G^0\left({\mathbf 0},{\mathbf r'};\omega\right),\label{eq:TmatrixGrennsfunctionsupplemental}
\end{equation}
with $G^0\left({\mathbf r},{\mathbf r'};\omega\right)$ the unperturbed Green's function and the $T$-matrix
\begin{equation}
    T\left(\omega\right)=\frac{U_0}{1-U_0g^0\left(\omega\right)}.\label{eq:Tmatrixsupplemental}
\end{equation}
In our case $U_0$ is given by the matrix $U_0 = v_F {\cal A}_x \sigma_x + v_F {\cal A}_y\sigma_y\tau_z$.
Notice $\left[U_0,\Delta_{\rm ivc}\sigma_{x}\tau_{x}\rho_{z}\right]=0$.

The quasiparticle spectrum may be recovered by finding the poles of $G$.
Thus, to calculate the perturbation-induced in-gap bound state spectrum it is sufficient to find the poles of the $T$-matrix. Thus, the in-gap states have energy $\omega$ such that the matrix
\begin{equation}
    Q\equiv 1-U_0g^0\left(\omega\right),\label{eq:Qpolesmatrix}
\end{equation}
has a zero eigenvalue.
The eigenvalues of $Q$ are
\begin{equation}
    \lambda_Q=1\pm v_F{\cal A}C\left(\Delta_{\rm ivc}\pm\omega\right),\label{eq:lambdaQsupplemental}
\end{equation}
with ${\cal A}=\sqrt{{\cal A}_x^2+{\cal A}_y^2}$.
Each of these four eigenvalues is doubly degenerate.
Notice the particle-hole symmetry. If $\omega^*$ makes one eigenvalue go to zero, then $-\omega^*$ makes another eigenvalue go to zero as well.

In order to make further progress, we must address $C$, which is not a constant, but rather a function of $\omega$ and $\Delta_{\rm ivc}$.

\subsection{Linear-in-$\epsilon$ density of states}
Using the linear density of states we employed in Eq.~\eqref{eq:DOSsupplementary}, we find
\begin{equation}
    C=\frac{1}{\left(W/2\right)^{2}}\log\left[1+\frac{\left(W/2\right)^{2}}{\Delta_{{\rm ivc}}^{2}-\omega^{2}}\right].\label{eq:ClinearDOSsupplemental}
\end{equation}
It is instructive to find the impurity strength at which the gap closes , i.e., $\omega=0$ leads to a zero $\lambda_Q$.
This is given by ${\cal A}_0$,
\begin{equation}
    \frac{v_{F}{\cal A}_{0}}{W/2}=\frac{W/2}{\Delta_{{\rm ivc}}}\frac{1}{\log\left[1+\left(\frac{W/2}{\Delta_{{\rm ivc}}}\right)^{2}\right]}.\label{eq:vfA0supplemental}
\end{equation}
Additionally, it is interesting to find the behavior of the in-gap state at very small strain, $v_{F}{\cal A}\ll W$.
In this limit, we find
\begin{equation}
    \omega\approx\pm\Delta_{{\rm ivc}}\left[1-\left(\frac{W/2}{2\Delta}\right)^{2}\exp\left(-\frac{W/2}{2\Delta}\frac{W/2}{v_{F}{\cal A}}\right)\right],\label{eq:smallstrainimpuritysupplemental}
\end{equation}
i.e., the bound state is exponentially close to the gap edge.
This behavior can indeed be seen in Fig.~\ref{fig:imuputiyspectrum} at small $\cal A$.

\begin{figure}
\begin{centering}
\includegraphics[scale=0.7]{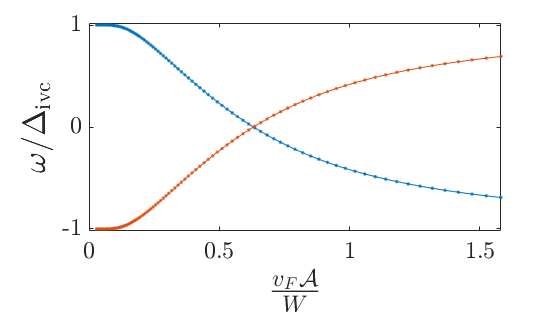}
\par\end{centering}
\caption{ \label{fig:imuputiyspectrum}
Bound state spectrum of the strain impurity problem. 
The two branches of the spectrum appear in blue and red, showing the particle-hole symmetry.
Each of the branches is doubly degenerate, reflecting the Kramers-like time-reversal symmetry ${\cal T}'$ of the Hamiltonian, even in the presence of the strain perturbation.
For this plot we used $\Delta_{\rm ivc}=W/5$.
}
\end{figure}

\subsection{Constant density of states}
In order to shed some light on the results of the last section, as well as to further facilitate the comparison to a magnetic impurity embedded in a singlet superconductor, let us repeat the calculation with a constant density of states, ${\cal N}\left(\epsilon\right)={\cal N}_0$. 
In this case,
\begin{equation}
    C=\frac{\pi{\cal N}_{0}}{\sqrt{\Delta_{{\rm ivc}}^{2}-\omega^{2}}}.\label{eq:CconstantDOSsupplemental}
\end{equation}
The bound state spectrum is thus given by
\begin{equation}
    \omega=\pm\Delta_{\rm ivc}\frac{1-s^2}{1+s^2},\label{eq:constantDOSspectrumsupplemental}
\end{equation}
with the dimensionless impurity strength $s=\pi{\cal N}_{0}v_{F}{\cal A}$.
It is worth mentioning that the result of Eq.~\eqref{eq:constantDOSspectrumsupplemental} is identical to the bound state energies of in-gap states induced in a singlet superconductor in the presence of a magnetic impurities, see, e.g., Ref.~\cite{TSCshibaChainsvonOppenImpurity}.
Thus, although the perturbation breaks no symmetries of the mean-field Hamiltonian, due to its matrix form which commutes with the order parameter -- an in-gap state (which may eventually reach all the way to zero energy) becomes possible.

\section{Details of the Abrikosov-Gorkov calculations}
Calculations in this section closely resemble the techniques used to treat magnetic impurities in singlet superconductors~\cite{AGmethodAbrikosov1959theory,maki1969gapless} and excitonic condensation in semiconductors~\cite{AGmethodZittartz} and graphene~\cite{AGmethodBistritzerMacdonald}.
We combine self-consistent mean-field calculations for the order parameter $\Delta_{\rm ivc}$, together with a self-consistent Born approximation, which allows us to handle the random strain fluctuations.

The key player in this treatment of the problem is the Green's function, which satisfies (within the mean-field approximation)
\begin{equation}
    G\left({\mathbf k},\omega\right) = 
    \left(
    i\omega-h_{\rm MF} - \Sigma_{\rm SCBA}
    \right)^{-1},\label{eq:Greensfunctionsupplemental}
\end{equation}
with the mean-field Hamiltonian
\begin{equation}
    h_{\rm MF} = u\left(k_x\sigma_x\tau_z+k_y\sigma_y\right)+\Delta_{\rm ivc} \sigma_x\tau_x\rho_z.\label{eq:hMFmeanfieldsupplemental}
\end{equation}


Within the self-consistent Born approximation, the self-energy is given by
\begin{equation}
    \Sigma_{\rm SCBA}\left({\mathbf k},\omega\right)=
    \left\langle
    \sum_{|\mathbf p| < \Lambda_c}
    {\cal U}_{{\mathbf k}-{\mathbf p}}
    G\left({\mathbf p},\omega\right)
    {\cal U}_{{\mathbf p}-{\mathbf k}}
    \right\rangle_{\rm disorder}
    \label{eq:SCBAmatrixsupplemental}
\end{equation}
where the matrix $\cal U$ represents the random strain perturbation in momentum space, $H_{\rm str} = \sum_{\mathbf{kq}}\Psi^{\dagger}_{\mathbf{k+q}}{\cal U}_\mathbf{q}\Psi_\mathbf{k}$,
and $\left\langle...\right\rangle_{\rm disorder}$ stands for averaging over disorder configurations.
Notice that Eq.~\eqref{eq:SCBAmatrixsupplemental} contains within it the full Green's function, such that Eq.~\eqref{eq:Greensfunctionsupplemental} and Eq.~\eqref{eq:SCBAmatrixsupplemental} must be solved together self-consistently.

The strain operator has the form
\begin{equation}
    {\cal U}_{\mathbf q} = u\left(
    A_{x,{\mathbf q}}\sigma_x
    +
    A_{y,{\mathbf q}}\sigma_y\tau_z
    \right).
\end{equation}
Assuming the disorder averaging eliminates any anisotropy introduced by a specific disorder configuration, we may assume
$\left\langle A_{x,{\mathbf q}}^* A_{y,{\mathbf q}}\right\rangle_{\rm disorder}
=
\left\langle A_{y,{\mathbf q}}^* A_{x,{\mathbf q}}\right\rangle_{\rm disorder}=0$,
and
$\left\langle A_{x,{\mathbf q}}^* A_{x,{\mathbf q}}-A_{y,{\mathbf q}}^* A_{y,{\mathbf q}}\right\rangle_{\rm disorder}=0$.
Using the fact that the real-space strain perturbation is real, we have $A_{i,-{\mathbf q}}=A^*_{i,{\mathbf q}}$.
Employing the ansatz,
\begin{equation}
    G_{\rm SCBA}\left({\mathbf k},\omega\right) = 
    \left(
    i\tilde{\omega}-uk_x\sigma_x\tau_z-uk_y\sigma_y-\tilde{\Delta}\sigma_x\tau_x\rho_z
    \right)^{-1},\label{eq:GreensfunctionSCBAsupplemental}
\end{equation}
the self-energy takes the form
\begin{equation}
    \Sigma_{\rm SCBA}\left({\mathbf k},\omega\right)-=
    \left\langle
    \sum_{\mathbf p}
    \left|u\vec{A}_{\mathbf{p-q}}\right|^2
    \frac{i\tilde{\omega}+\tilde{\Delta} \sigma_x\tau_x\rho_z}{\epsilon_{\mathbf p}^2 +\tilde{\Delta}^2+\tilde{\omega}^2}
    \right\rangle_{\rm disorder}.
    \label{eq:SCBAselfenergydetailedsupplemental}
\end{equation}

For simplicity, we now make the standard assumption that the disorder potential $\vec{A}_{\mathbf {p-q}}$ depends mostly on the relative angle between the initial and final momenta, i.e., $\vec{A}_{\mathbf {p-q}}\approx\vec{A}_\theta$, where the angle is defined by $\tan\theta=\frac{p_y-q_y}{p_x-q_x}$.
This assumption is due to the singularity of the Green's function at the Fermi level, such that the main contribution in the sum appearing in Eq.~\eqref{eq:SCBAselfenergydetailedsupplemental} is by momenta with $\left|{\mathbf p}\right|\approx p_F$, with $p_F$ the Fermi momentum.
With this approximation, we find the self-consistent equations for $\tilde{\omega}$ and $\tilde{\Delta}$,
\begin{equation}
    \tilde{\omega} = \omega + \tilde{\omega} \Gamma \frac{W}{2}
    \int d\epsilon\frac{{\cal N}\left(\epsilon\right)}{\epsilon^2+\tilde{\Delta}^2+\tilde{\omega}^2},\label{eq:omegatildesupplemental}
\end{equation}
\begin{equation}
    \tilde{\Delta} = \Delta - \tilde{\Delta} \Gamma \frac{W}{2}
    \int d\epsilon\frac{{\cal N}\left(\epsilon\right)}{\epsilon^2+\tilde{\Delta}^2+\tilde{\omega}^2},\label{eq:Deltatildesupplemental}
\end{equation}
where ${\cal N}\left(\epsilon\right)$ is the density of states per moir\'e unit-cell $\Omega_{u.c.}$, $W$ is the bandwidth, and the scattering rate $\Gamma$ is
\begin{equation}
    \Gamma = \frac{2\Omega}{W \Omega_{u.c.}}\left\langle \int d\theta 
    u^2 \vec{A}_{\theta}^\dagger \cdot\vec{A}_{\theta}
    \right\rangle_{\rm disorder}\label{eq:Gammasupplemental}.
\end{equation}

Finally, to complete the picture, we must include the self-consistent calculation of the order parameter itself, with the gap equation
\begin{align}
    \Delta_{{\rm ivc}}&=-2\frac{U}{\beta\Omega}\sum_{\omega\mathbf{k}}{\rm Tr}\left\{ \sigma_{x}\tau_{x}\rho_{z}G_{\rm SCBA}\left(\mathbf{k},\omega\right)\right\}\nonumber\\
    &=2U \frac{1}{\beta}\sum_\omega\int d\epsilon\frac{{\cal N}\left(\epsilon\right)}{\epsilon^{2}+\tilde{\Delta}^{2}+\tilde{\omega}^{2}}\tilde{\Delta}.\label{eq:selfconsistentgapsupplemental}
\end{align}
Equations~\eqref{eq:omegatildesupplemental},\eqref{eq:Deltatildesupplemental}, and ~\eqref{eq:selfconsistentgapsupplemental} form a set of closed equations that can be solved together to recover $\Delta_{\rm ivc}$ and $G_{\rm SCBA}$ in the presence of strain fluctuations.

Lastly, we may also calculate the tunneling density of states, given by
\begin{align}
    \tilde{\cal N} \left(\epsilon\right) &= \frac{1}{\pi}{\rm Im} \frac{1}{\Omega}\sum_{\mathbf k} {\rm Tr} G_{\rm SCBA}\left({\mathbf k},\omega\to i\epsilon\right)\nonumber\\
    & = \frac{1}{\pi}{\rm Im}\int d\epsilon{\cal N}\left(\epsilon\right)\frac{i\tilde{\omega}}{\epsilon^{2}+\tilde{\Delta}^{2}+\tilde{\omega}^{2}}|_{\omega\to i\epsilon}.\label{eq:tunnelingdossupplemental}
\end{align}
When $\Delta_{\rm ivc}$ is finite, one expects to find a gap in $\tilde{\cal N} \left(\epsilon\right)$ near zero energy.
We denote the value of this gap as $\omega_g$.

\subsection{Estimating the disorder scale $\Gamma$}

The disorder parameter $\Gamma$ determines completely the properties of the disorder-averaged Green's function within our SCBA treatment. Given the homostrain tensor $\epsilon^{\rm homo}_{\alpha\beta}$ introduced in Eq.~\eqref{eq:epsilonhomosupp},
the effective pseudo-gauge field can be expressed as~\cite{KoshinoStrain}

\begin{equation}
    {\mathbfcal A}=\frac{\sqrt{3}\beta}{a}
    \left(\frac{\epsilon^{\rm homo}_{xx}-\epsilon^{\rm homo}_{yy}}{2},-\epsilon^{\rm homo}_{xy}\right),\label{eq:gaugefromstrainsupplemental}
\end{equation}
where $a=0.246\,{\rm nm}$ is the monolayer graphene lattice constant, and $\beta\approx3.14$ is a dimensionless parameter of graphene, which quantifies the sensitivity of nearest-neighbor hopping strength to a change in their relative distance.

As an example, let us consider a scenario where root-mean-square strain is $\epsilon_{\rm RMS}$, approximate $u\approx 0.1 v_F$~\cite{CaoCorrelatedInsulator}, and use $\hbar v_F/a \approx 2.68 eV$,
such that
\begin{equation}
    u\left|{\cal A}_{\rm RMS}\right|\approx 14.5\,{\rm meV}\times\epsilon_{\rm RMS}\left[\%\right].\label{eq:vfAsupplemental}
\end{equation}
In order to continue, we examine the specific example of Gaussian correlated disorder, i.e., assume the following correlations,
\begin{equation}
    \left\langle
    {\cal A}_i \left(\mathbf{r}\right)
    {\cal A}_j \left(\mathbf{r'}\right)
    \right\rangle_{\rm disorder}
    =\delta_{ij} {\cal A}_{\rm RMS}^2 \exp\left({-\frac{\left|\mathbf{r}-\mathbf{r'}\right|^2}{2\xi_{\rm dis.}^2}}\right),
\end{equation}
where $\xi_{\rm dis.}$ is the disorder correlation length. 
Following a straightforward Fourier transform, we find
\begin{equation}
    \left\langle
    {A}_i \left(\mathbf{q}\right)
    {A}_j \left(\mathbf{-q}\right)
    \right\rangle_{\rm disorder}
    =\delta_{ij}{\cal A}_{\rm RMS}^2
    \frac{1}{\Omega}\xi_{\rm dis.}^2\exp\left(-\frac{\left|\mathbf {q}\right|^2\xi_{\rm dis.}^2}{2}\right).
\end{equation}
Since we began with the expression in Eq.~\eqref{eq:SCBAselfenergydetailedsupplemental}, our main focus is on momenta near the Fermi level. We thus replace $\mathbf{q}\approx k_F=0$ in our estimates below.
Estimating the MATBG bandwidth as $W\approx 20\,{\rm meV}$,
and using Eq.~\eqref{eq:Gammasupplemental} we find
\begin{equation}
    \Gamma /W
    \approx
    \frac{4\xi_{{\rm dis.}}^{2}}{A_{{\rm u.c.}}}\left(\frac{u{\cal A}_{{\rm RMS}}}{W}\right)^{2}
    \approx 2.1 \times \left(\epsilon_{\rm RMS}\left[\%\right] 
    \times
    \xi_{\rm dis.} \left[ a_{\rm m} \right ]\right)^2\label{eq:GammaEstimationSupplemetal},
\end{equation}
where $a_{\rm m}\approx\sqrt{A_{\rm u.c.}}$ is the moir\'e lattice constant.

For example, for $\epsilon_{\rm RMS}=0.1\%$ and $\xi_{\rm dis.}=3 a_{\rm m}$ we find $\Gamma/W\approx 0.19$.
We can thus conclude that for appreciable yet reasonable homostrain disorder, $\Gamma$ may indeed be large enough to diminish both $\Delta_{\rm ivc}$ and $\omega_g$ significantly, and possibly completely close the gap in the TDOS.

\subsection{Critical disorder strength}
If the energy scale associated with the random strain fluctuations $\Gamma$ is sufficiently large, the K-IVC order may be completely washed away, even at zero temperature.
Using the linear density of states approximation, we can recover this critical $\Gamma_c$, by setting $\Delta_{\rm ivc},\tilde{\Delta}\to0$ and $T=0$.
We note that it is important not to set $\Delta_{\rm ivc}=\tilde{\Delta}$ exactly, although both quantities are taken to zero. This is because in the gap equation we must divide both sides by $\Delta_{\rm ivc}$
Under these assumptions,
\begin{equation}
    \tilde{\omega}=\omega+\frac{\Gamma_c}{W/2}\log\left(1+\frac{\left(W/2\right)^{2}}{\tilde{\omega}^{2}}\right)\tilde{\omega},\label{eq:tildeomegaCriticalGammasupplemental}
\end{equation}
\begin{equation}
    \tilde{\Delta}=\frac{\Delta}{1+\frac{\Gamma_{c}}{W/2}\log\left(1+\frac{\left(W/2\right)^{2}}{\tilde{\omega}^{2}}\right)},\label{eq:tildedeltaCriticalGammasupplemental}
\end{equation}
from which we also calculate
\begin{equation}
    d\omega = d\tilde{\omega}\left[1-\frac{\Gamma_c}{W/2}\log\left(1+\frac{\left(W/2\right)^{2}}{\tilde{\omega}^{2}}\right)+2\Gamma_c\frac{W/2}{\tilde{\omega}^{2}+\left(W/2\right)^{2}}\right].\label{eq:domegadomegatildesupplemental}
\end{equation}
Plugging this relation into the zero temperature gap equation and changing variables, we find
\begin{equation}
    \frac{U_{c}}{U}=\int_{-\infty}^{\infty}\frac{dx}{2\pi}\frac{1-g\log\left(1+\frac{1}{x^{2}}\right)+2g\frac{1}{x^{2}+1}}{1+g\log\left(1+\frac{1}{x^{2}}\right)}\log\left(1+\frac{1}{x^{2}}\right),\label{eq:criticalgammagapequation}
\end{equation}
where $g=2\Gamma_c/W$. The self-consistent Eq.~\eqref{eq:criticalgammagapequation} relates the interaction strength $U$ to the critical disorder strength via a dimensionless integral.
For relatively weak disorder, $g\ll 1$, we may expand Eq.~\eqref{eq:criticalgammagapequation} to linear order in $g$, to find the relation
\begin{equation}
    \Gamma =\frac{U_c}{\log 8}\left(1-\frac{U_c}{U}\right),
\end{equation}
which is valid close for $U \approx U_c$, such that $g \ll 1$.
\subsection{Gapless K-IVC}
Unlike in the translation-symmetry invariant case, the gap in the tunneling density of states, $2\omega_g$, does not necessarily equal the self-consistent order parameter $\Delta_{\rm ivc}$.
More concretely, there exists a regime in which $\Delta_{\rm ivc}$ is finite, yet $\omega_g=0$ and the tunneling density of states is gapless.

Let us find the disorder strength $\Gamma_g$ for which this gap closes. To do so, we calculate the tunneling density of states at zero energy $\tilde{\cal N}\left(\epsilon=0\right)$. Plugging $\omega\to 0$ in the expressions for $\tilde{\omega}$ and $\tilde{\Delta}$, we find
\begin{equation}
        \tilde{\omega}^{2}=\frac{\left(W/2\right)^{2}}{\exp{\frac{W/2}{\Gamma}}-1}-\left(\Delta_{{\rm ivc}}/2\right)^{2}.
\end{equation}
The tunneling density of states is then given by
\begin{equation}
    {\cal \tilde{N}}\left(\epsilon=0\right)=\frac{1}{2\pi}\frac{1}{\Gamma W}\times{\rm Re}\left\{ \tilde{\omega}_{\omega\to0}\right\}. 
\end{equation}
Thus, the gap closes, i.e., there is a finite value of $\tilde{\cal N}$ at zero energy, for
\begin{equation}
    \Gamma\geq\Gamma_g=\frac{W/2}{\log\left(1+\frac{W^{2}}{\Delta_{{\rm ivc}}^{2}}\right)}.\label{eq:GammaGapclosesupplemental}
\end{equation}
Notice Eq.~\eqref{eq:GammaGapclosesupplemental} is not a closed-form expression for $\Gamma_g$, since in the right hand side $\Delta_{\rm ivc}$ depends on $\Gamma$ itself.

\begin{figure}
\begin{centering}
\includegraphics[scale=0.65]{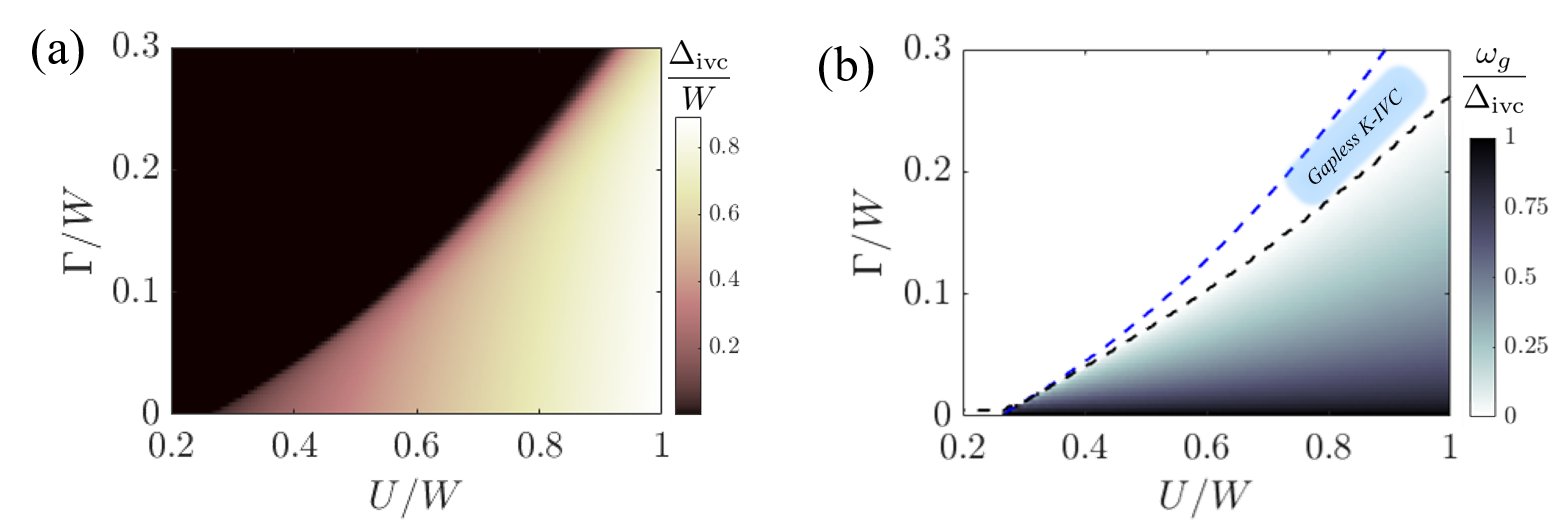}
\par\end{centering}
\caption{ \label{fig:mapsofU}
(a) 
The order parameter $\Delta_{\rm ivc}$ as a function of interaction strength and the disorder energy scale $\Gamma$.
At a given interaction strength $U$, a critical $\Gamma$ exists at which the order parameter vanishes.
(b)
The ratio of the gap in the TDOS, $\omega_g$, to $\Delta_{\rm ivc}$.
As $\Gamma$ increases, $\omega_g$ deviates from $\Delta_{\rm ivc}$ appreciably.
Moreover, there exists a gapless K-IVC region bounded by the vanishing of the gap (black dashed line) and of the order parameter (blue dashed line).
Calculations were done at zero temperature, using the linear DOS ${\cal N}_{\rm lin.}$.
}
\end{figure}

In the main text, we have shown that the gapless K-IVC regime exists over a range of temperatures and disorder strengths. For completeness, we also show how the gapless K-IVC at zero temperature evolves as a function of the repulsive interaction strength $U$.
This is plotted in Fig.~\ref{fig:mapsofU}.

\subsection{Conductivity calculations}
The primary method by which the K-IVC gap is extracted in experiments is by derivation of the activation energy from transport measurements.
Namely, the longitudinal conductivity $\sigma_{xx}$ is measured as a function of temperature at the appropriate filling corresponding to the insulating phase, and fitted to 
$\propto \exp \left(-E_{\rm act.}/T\right)$, where the activation energy $E_{\rm act.}$ presumably coincides with the K-IVC order parameter $\Delta_{\rm ivc}$.

In order to put our results thus far in a similar context, we employ the Kubo-Bastin formula for the dc conductivity~\cite{KuboBASTIN19711811},
\begin{equation}
    \sigma_{xx}\propto\int d\epsilon\left(-\frac{df}{d\epsilon}\right) {\cal S}\left(\epsilon\right),\label{eq:sigxxbastinsupplemental}
\end{equation}
where $f$ is the Fermi function, and the kernel of the integral is
\begin{equation}
    {\cal S}\left(\epsilon\right)
    =
    \frac{1}{\Omega}\sum_{\mathbf{k}}
    {\rm Tr}\left\{
    \left\langle
    j_x
    {\rm Im}G_{\rm SCBA}\left({\mathbf k},\omega\to i\epsilon\right)
    j_x
    {\rm Im}G_{\rm SCBA}\left({\mathbf k},\omega\to i\epsilon\right)
    \right\rangle_{\rm disorder}\right\},\label{eq:Sepsilonsupplemental}
\end{equation}
where the appropriate current operator in our case is $j_x=\partial_{k_x} h_{\rm MF} =u \sigma_x\tau_z$.

The full calculation of Eq.~\eqref{eq:Sepsilonsupplemental}, including the so-called vertex corrections to the current operator, are beyond the scope of this work.
However, qualitatively most of the disorder effects stemming from the Abrikosov-Gor'kov treatment we perform may be captured by approximating
\begin{equation}
    {\cal S}\left(\epsilon\right)
    \approx
    \frac{1}{\Omega}\sum_{\mathbf{k}}
    {\rm Tr}\left\{
    j_x
    {\rm Im}G_{\rm SCBA}\left({\mathbf k},\omega\to i\epsilon\right)
    j_x
    {\rm Im}G_{\rm SCBA}\left({\mathbf k},\omega\to i\epsilon\right)
    \right\},\label{eq:Sepsilonnovertexcorrsupplemental}
\end{equation}
where we emphasize that the Green's function $G_{\rm SCBA}$ is calculated self-consistently accounting for the strain disorder.

Plugging in the Green's function [Eq.~\eqref{eq:GreensfunctionSCBAsupplemental}], we find by a straightforward calculation that the conductivity kernel ${\cal S}\left(\epsilon\right)$ has the following form,
\begin{align}
    {\cal S}\left(\epsilon\right)
    &\approx
    \frac{1}{\Omega}\sum_{\mathbf{k}}
    \frac{{\rm Re}\left(\tilde{\omega}\right)^{2}-{\rm Im}\left(\tilde{\Delta}\right)^{2}}{\left(\epsilon_{\mathbf{k}}^{2}+\left|\tilde{\omega}\right|^{2}+\left|\tilde{\Delta}\right|^{2}\right)^{2}-4\left[{\rm Im}\left(\tilde{\omega}\tilde{\Delta}\right)\right]^{2}-4\left[{\rm Im}\left(\tilde{\omega}\right)\right]^{2}\epsilon_{\mathbf{k}}^{2}}
    \times\nonumber\\
    &\times \left\{ 1+\frac{4\left[{\rm Im}\left(\tilde{\omega}\right)\right]^{2}\epsilon_{\mathbf{k}}^{2}}{\left(\epsilon_{\mathbf{k}}^{2}+\left|\tilde{\omega}\right|^{2}+\left|\tilde{\Delta}\right|^{2}\right)^{2}-4\left[{\rm Im}\left(\tilde{\omega}\tilde{\Delta}\right)\right]^{2}-4\left[{\rm Im}\left(\tilde{\omega}\right)\right]^{2}\epsilon_{\mathbf{k}}^{2}}\right\}. 
\end{align}
Notice that $\tilde{\omega}$ and $\tilde{\Delta}$ are generally both \textit{complex} numbers and functions of energy $\epsilon$, through the analytic continuation $\omega\to i\epsilon$, preceding the iterative solution of Eqs.~\eqref{eq:omegatildesupplemental}--\eqref{eq:Deltatildesupplemental}.

Importantly we find that the functions 
${\rm Re}\tilde{\omega}$ and ${\rm Im}\tilde{\Delta}$ both vanish for $\left|\epsilon\right|<\omega_g$, i.e., the conductivity kernel ${\cal S} \left(\epsilon\right)$ has a gap of the size $2\omega_g$.
This fact, in conjunction with the relation in Eq.~\eqref{eq:sigxxbastinsupplemental}, explains our observation given in the main text, that the activation energy actually coincides with $\omega_g$.
To see why, let us approximate ${\cal S} \left(\epsilon\right)\approx {\cal S}_0 \Theta\left(\left|\epsilon\right|-\omega_g\right)$ ($\Theta$ is the Heaviside function), such that
\begin{equation}
    \sigma_{xx}\propto{\cal S}_0 \int_{\omega_g}^\infty d\epsilon\left(-\frac{df}{d\epsilon}\right)
    =\int_{\omega_g}^\infty d\epsilon\frac{{\cal S}_0}{4T\cosh^2\frac{\epsilon}{2T}}
    =\frac{{\cal S}_0}{2}\left(1-\tanh\frac{\omega_g}{2T}\right)
    =\frac{{\cal S}_0}{1+e^{\frac{\omega_g}{T}}}
    ,\label{eq:sigxxactivationsupplemental}
\end{equation}
which for sufficiently low temperatures $T\ll\omega_g$, leads to the $\propto \exp \left(-\omega_g/T\right)$ activated behavior.

\clearpage
\let\vaccent=\v 
\renewcommand{\v}[1]{\ensuremath{\mathbf{#1}}} 

\end{widetext}

\end{document}